\documentclass[
    ,final            % use final for the camera ready runs
  ]
  {aipproc}

\layoutstyle{8x11single}
\usepackage{amsmath}
\usepackage{amsbsy}
\usepackage{amssymb}
\usepackage{subfigure}
\begin{document}

\title{Universal scaling form of AC response in variable range hopping}


\author{Joakim Bergli}{
  address={Physics Department, University of Oslo 
P.O.Box 1048 Blindern, 0316 Oslo, Norway}
}

\author{Yuri M. Galperin}{
  address={Physics Department, University of Oslo 
P.O.Box 1048 Blindern, 0316 Oslo, Norway}
}

\begin{abstract}
  We have studied the AC response of a hopping model in the variable
  range hopping regime by dynamical Monte Carlo simulations. We find
  that the conductivity as function of frequency follows a universal
  scaling law. We also compare the numerical results to various
  theoretical predictions. Finally, we study the form of the
  conducting network as function of frequency.
\end{abstract}

\maketitle

%%%%%%%%%%%%%%%%%%%%%%%%%%%%%%%%%%%%%%%%%%%%
%% MAINMATTER
%%%%%%%%%%%%%%%%%%%%%%%%%%%%%%%%%%%%%%%%%%%%

\section{Introduction}

There are many examples of disordered solids that display a universal
scaling form of the AC conductance~\cite{dyre}. This is seen if one
considers the conductivity $\sigma(\omega)$ as function of frequency
$\omega$ at different temperatures. It is found that the curves for
different temperatures collapse on a single universal curve if scaled
according to the following equation:
\begin{equation}\label{eq1}
  \frac{\sigma(\omega)}{\sigma(0)} 
  = F\left(\frac{\omega}{T\sigma(0)}\right)
\end{equation}
Such a scaling implies that the only relevant timescale in the system
is $1/T\sigma(0)$.

Several models displaying this behavior have been studied~\cite{dyre},
including hopping models. However, electron-electron interaction was
neglected and
only the regime of nearest neighbor hopping was considered.
It is therefore interesting to see if an interacting system displaying
variable range hopping (VRH) will possess the same universality property.

\section{Model}

We have studied the standard 2D lattice model of hopping transport
\cite{davies,ES}. It consist of a square lattice of $L\times L$ sites. 
 Each of the sites can
be either empty or occupied (double occupancy is not allowed).
The simulations were performed for the filling factor of $\nu=1/2$, so that
the number of electrons is half the number of sites. Disorder is
introduced by assigning a random energy in the range $[-U,U]$ to
each site (in our numerics we have used $U = e^2/d$ where $d$ is the
lattice constant). Each site is also given a compensating charge $\nu e$
so that the overall system is charge neutral. The charges interact
via the Coulomb interaction. In the following the unit of energy is
chosen as the Coulomb energy of unit charges on nearest neighbor
sites, $e^2/d$. The unit of temperature is then $e^2/dk$.
It is known that this model shows VRH DC
conduction according to the Efros-Shklovskii law, $\sigma(0) \sim
e^{-(T_0/T)^{1/2}}$, in a rather broad temperature
range~\cite{tsigankov,bergli}.

To simulate the time evolution we used the dynamic Monte Carlo method
introduced in Ref.~\cite{tsigankov} for simulating DC transport.     
The only difference is that we apply an AC electric field            
$E=E_0\cos\omega t$ and that, following Ref.~\cite{tenelsen},        
we use for the transition of an electron from site $i$ to site $j$ the
formula                                                               
\begin{equation}
\Gamma_{ij} = {\tau_0}^{-1}e^{-2 r_{ij}/\xi}\min                      
  \left(e^{-\Delta E/T},1\right) \label{eq:01}
\end{equation}
where $\Delta E$ is the energy of the phonon and $r_{ij}$ is the
distance between the sites.  $\tau_0$ contains material dependent and
energy dependent factors, which we approximate by their average value;
we consider it as constant and its value, of the order of $10^{-12}$
s, is chosen as our unit of time. Consequently, $\omega$ is measured
in units of $\tau_0^{-1}$ while the electric field is measured in
units of $e/d^2$. The algorithm works by starting from one
configuration of electrons on the sites and then a random transition
is chosen, weighted by the distance dependent part of the transition
rate. Then the jump is accepted or rejected based on the energy
dependent part of the transition rate. To each jump there is also
associated a time, $\Delta t$, depending on the number of rejected
trials before one is accepted~\cite{tsigankov2}. This generates a random
walk in the configuration space with statistical properties identical
to the real time evolution of the system.
%, where there naturally is no
%such strict time ordering of the jumps, but jumps can take place
%independently and simultaneously in different parts of the sample.

To extract the conductivity from the contributions of the
individual jumps we do as follows.  The current
density is $j = E_0(\sigma\cos\omega t+ \sigma_I\sin\omega t)$ which
means that ignoring oscillating terms we will have

\begin{equation}\label{eq:JR}
 J_R = \int_0^tdt j(t)\cos\omega t =  \frac{1}{2}E_0\sigma t\qquad
 J_I = \int_0^tdt j(t)\sin\omega t =  \frac{1}{2}E_0\sigma_It 
\end{equation}
The program will tell which jump took place and what time it took. Let
$\Delta x_i$ be the distance jumped in the direction of the field, and
$\Delta t_i$ be the time of jump number $i$. Then the current density
is $j_i = \frac{\Delta x_i}{\Delta t_i}$ and we replace the integral
\eqref{eq:JR} with the sum

\begin{equation}\label{JRs}
 J_R = \sum_i \Delta x_i \cos\omega t_i
\end{equation}
where $t_i$ is the time at which jump $i$ took place. From this we get the conductivity $\sigma=\frac{2}{E_0t}J_R$.

\section{Numerical confirmation of scaling}

We have simulated the response to an AC electric field as function of
frequency for different temperatures in the VRH regime (parts of the
numerical data were already presented in Ref. \cite{drichko}). In most
cases we used $L=100$, which is sufficient for the DC conductivity to
be well defined and sample independent. Only at frequencies $\omega>1$
was it necessary to use larger samples as the number of electrons that
will jump during one period of the field becomes small and the data
are noisy if too small samples are used. We compared $L=300$ and
$L=1000$ and found no significant difference, indicating that $L=300$
is sufficient.  In all cases we have chosen $E=T/10$ which is in the
upper range of ohmic response and therefore gives a large signal to
noise ratio.

\begin{figure}
\includegraphics[width=0.5\textwidth]{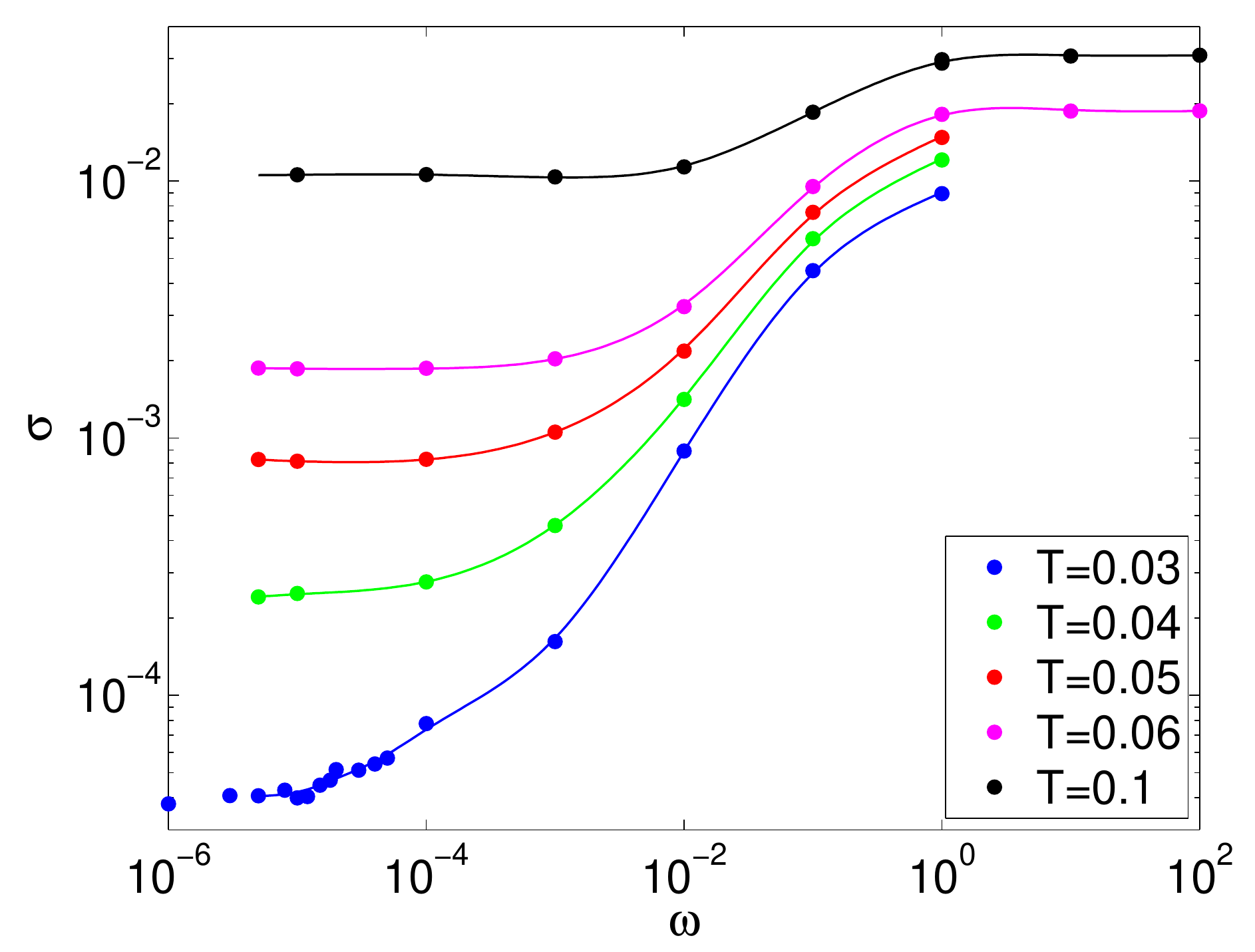}
\includegraphics[width=0.5\textwidth]{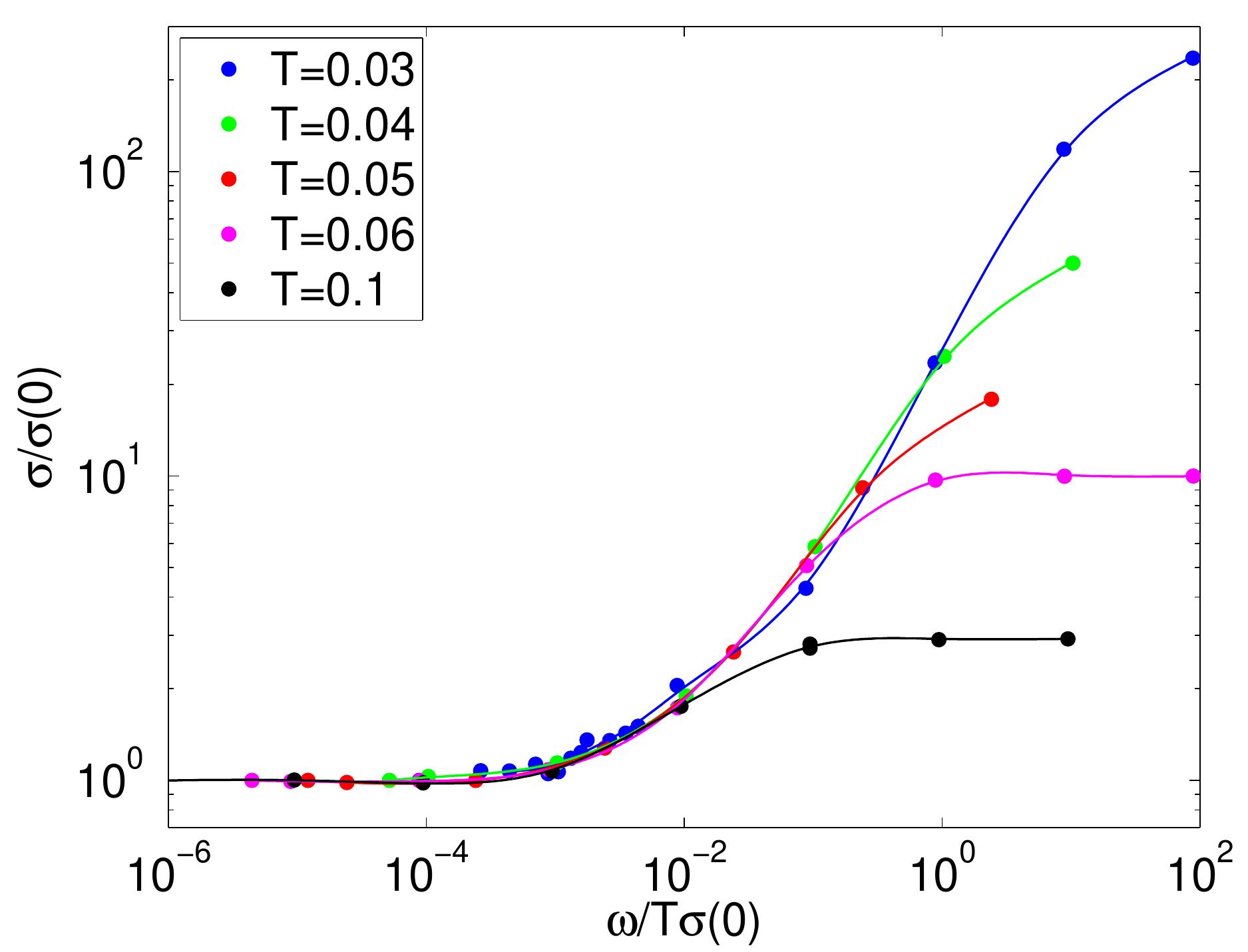}
\caption{\label{fig:condACScaling} Left:Real part of AC conductivity
  as function of frequency. Right: The same data rescaled according to
  Eq. \eqref{eq1}.  }
\end{figure}

The results are shown in Fig.~\ref{fig:condACScaling} (left).  The
general trend is the same at all temperatures: at low frequencies, the
conductivity is frequency independent. At some frequency it will start
to increase, and then saturate at high frequencies (for all
temperatures this happens around $\omega=1$. Looking at Eq
\eqref{eq:01} and recalling that $\tau_0 $ is our unit of time, we
understand that this happens when the frequency is higher thatn the
highest rate in the system).  To check the scaling form,
Eq. (\ref{eq1}), we replot the same data in
Fig. \ref{fig:condACScaling} (right), which shows
$\sigma(\omega)/\sigma(0)$ as function of $\omega/T\sigma(0)$.  As can
be seen, the curves fall on a common universal curve up to the point
where they show a trend to saturation, $\omega \lesssim
(10^{-2}-10^{-1}) T\sigma (0)$ . Note that in this region the
universal curve describes the values of $\sigma(\omega)$ differing by
factor of $\sim 300$.

We see that we numerically confirm the scaling equation \eqref{eq1},
but it is also interesting to see if the shape of the graph agrees
with theoretical predictions. Many different predictions have been
made for the form of $\sigma(\omega)$ for different models and in
various frequency regimes, see Refs.~\cite{dyre,BB,zvyagin,hunt} and
references therein. Some of these predictions conform to the scaling
form \eqref{eq1} either exactly or approximately, and all rely on some
approximate solution to the different models. We do not want to enter
into a full description of the different theoretical graphs, but
broadly speaking they fall into two groups. Those that rely on some
type of effective or homogeneous medium approximation and those based
on percolation theory.  There seems to be general agreement on the
overall physical picture. In the DC limit, there is the percolation
cluster on which the current flows, and with the critical resistors
determining the overall conductivity. As the frequency increases, the
cluster breaks into smaller regions, and also clusters not on the
percolation cluster starts to contribute to the conductivity. In the
high frequency limit, only isolated pairs of sites with high rates are
able to follow the oscillating field, and all traces of the
percolation cluster are lost. It is therefore to be expected that
effective medium approximations and expansions in small clusters will
work better at high frequencies while percolation based methods are
appropriate at low frequencies.

\begin{figure}
  \includegraphics[width=7.2cm]{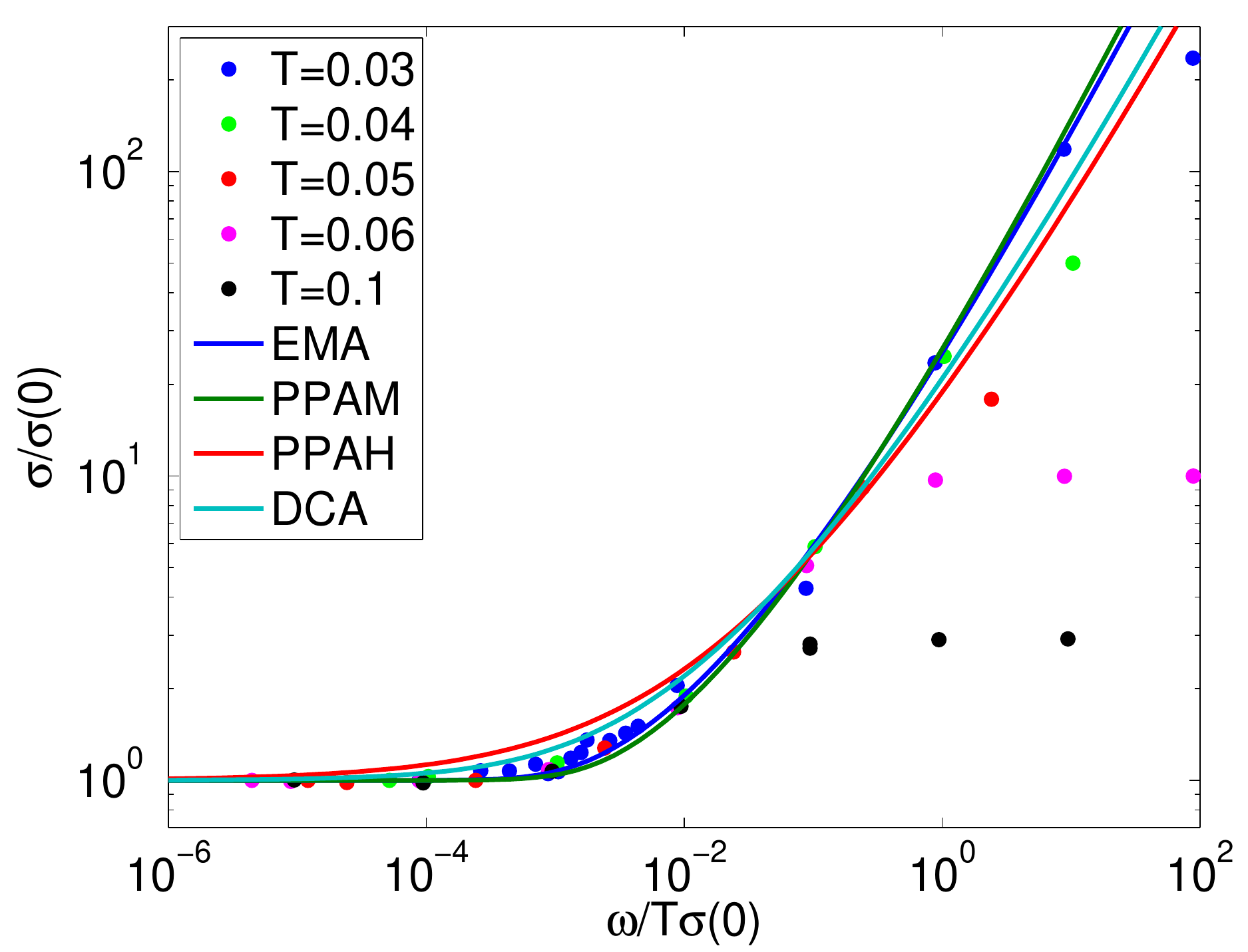}
\caption{\label{fig:condACScalingMedDyre}Real part of AC conductivity
  as function of frequency with best fits for each model. The four
  models are: Effective Approximation (EMA), Eq. (13) of \cite{dyre};
  Percolation Path Approximation for a macroscopic model (PPAM)),
  Eq. (25) of \cite{dyre}; Percolation Path Approximation for a
  hopping model (PPAH), Eq. (26) of \cite{dyre}; Diffusion Cluster
  Approximation (DCA), Eq. (28) of \cite{dyre} }
\end{figure}

We have fitted our numerical data to four different predictions for
the universal graph discussed in Ref.~\cite{dyre}. These are given by
their eqs. (13), (25), (26) and (28). Each function has an
undetermined scaling of the frequency which has to be fitted to the
numerical points. That is, they are given as functions of
$\alpha\frac{\omega}{T\sigma(0)}$ where $\alpha$ is an undetermined
parameter. Figure \ref {fig:condACScalingMedDyre} shows our numerical
data together with the best fits to each model. The models were fitted
as least squares fits to the logarithmic values of the $T=0.03$
points, excluding $\omega=1$ as this is where the curve saturates and
deviates from the universal curve. Note that the DCA (Diffusion
Cluster Approximation) in principle has another free parameter, the
fractal dimension of some poorly defined ``diffusion cluster''. We
have taken it as 1.35 as this was the best fitting value found from
the numerics in Ref. \cite{dyre}. By fitting to our data using this as
an additional parameter we would very likely improve the fit.

As we can see, all the curves are reasonably close to the numerical
data, even if some can be argued as better than others. It is
interesting to note that the two best fitting curves are the Effective
Medium Approximation (EMA) and the Percolation Path Approximation
(PPAM) which are at the opposite extremes in the set of models. The
first assumes a homogeneous effective medium, whole the second assumes
that all transport at all frequencies takes place an the DC
percolation cluster and that other regions never make a significant
contribution to the current. 

\section{Activity maps}

It is interesting to study the size and shape of conducting clusters
as the frequency changes. In the DC limit there is the percolation
network. As the frequency increases, the percolation cluster should break
up into smaller regions, and areas not on the percolation cluster
should start to contribute.

We will find the links which give large contributions
to the current.  For each link $l$ (that is, a connection between two
specific sites in the system), we sum all the contributions to the sum
\eqref{JRs} which comes from this link, let us call it $s_l$. This
includes jumps in both directions, and in the DC case, jumps in one
direction are positive and in the other negative. For AC it depends on
the direction of the field at the time a jump is made. We find that
most sites have at least some activity, and there are only a few sites
which are completely frozen. Many of the links will only have a small
number of jumps associated with it, or many jumps which are mainly
thermally driven, and not related to the electric field. They will
then give a small contribution to the sum \eqref{JRs}.  We want to
exclude those links which do not give a large contribution to the sum
\eqref{JRs} which can be written $J_R = \sum_l s_l$ where the sum is
over all links in the system.  Let $S$ be a threshold and ignore all
links which have $|s_l|\leq S$. This means that we keep all that have
a large absolute value, both positive and negative. Let $L^S$ be the
set of links that are kept with a particular value of $S$, and define
\begin{equation}
 J_R^S =\sum_{l\in L^S}s_l
\end{equation}
We specify a number $D\in[0,1]$ and find $S(D)$ such that
$J_R^S=DJ_R$. This means that we keep links such that the sum of the
contributions from those links is a fraction $D$ of the total.  We
find that $D=0.8$ gives a good balance between reducing the number of
active links while still keeping the important ones. In the following
we will show results at $D=0.8$, but the exact value should not affect
the results too strongly.

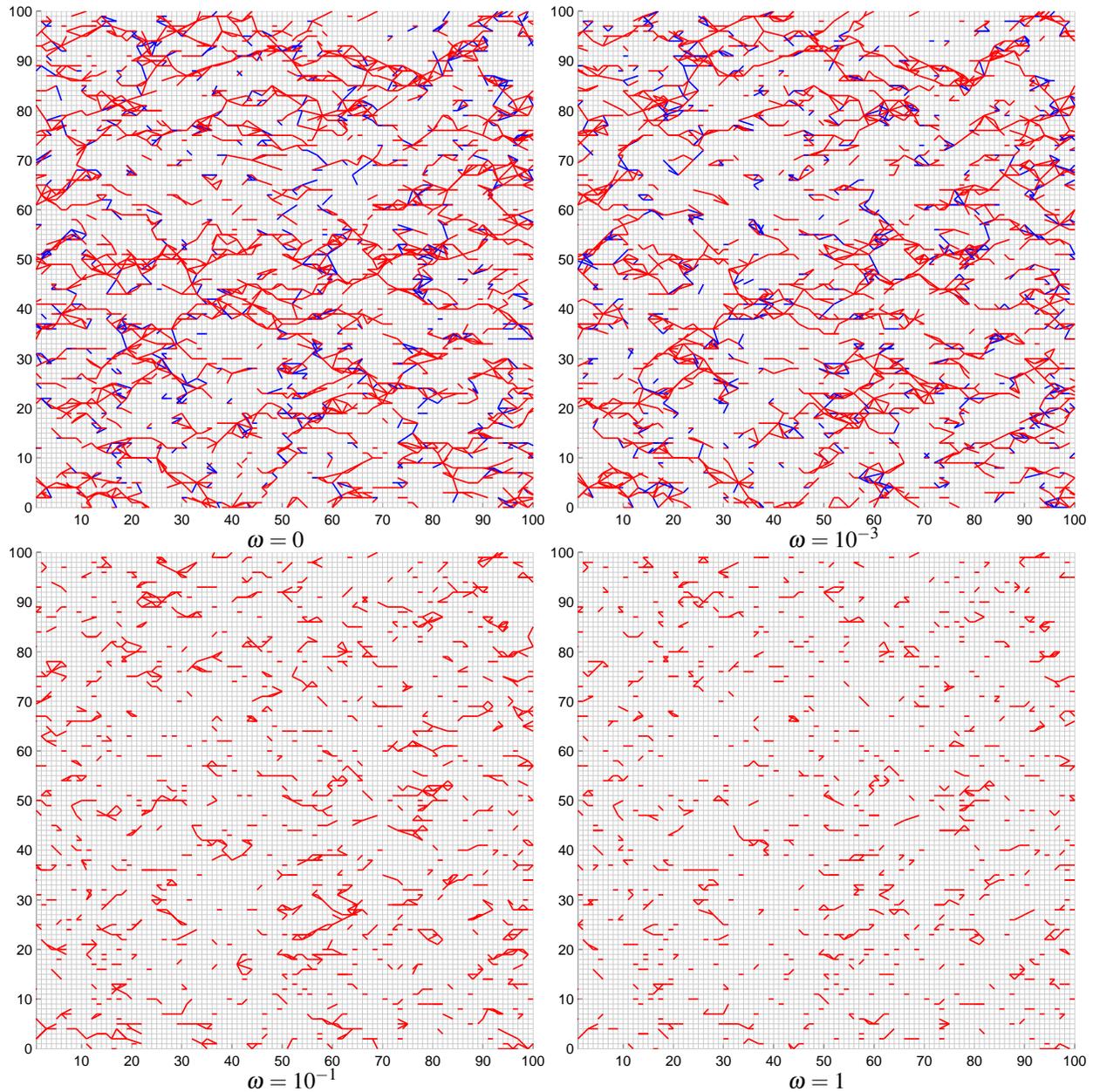
\begin{figure}
\setlength{\unitlength}{\textwidth}
\begin{picture}(1,1)(0,0)
\put(0,0.52){\includegraphics[width=0.5\unitlength,angle=0]{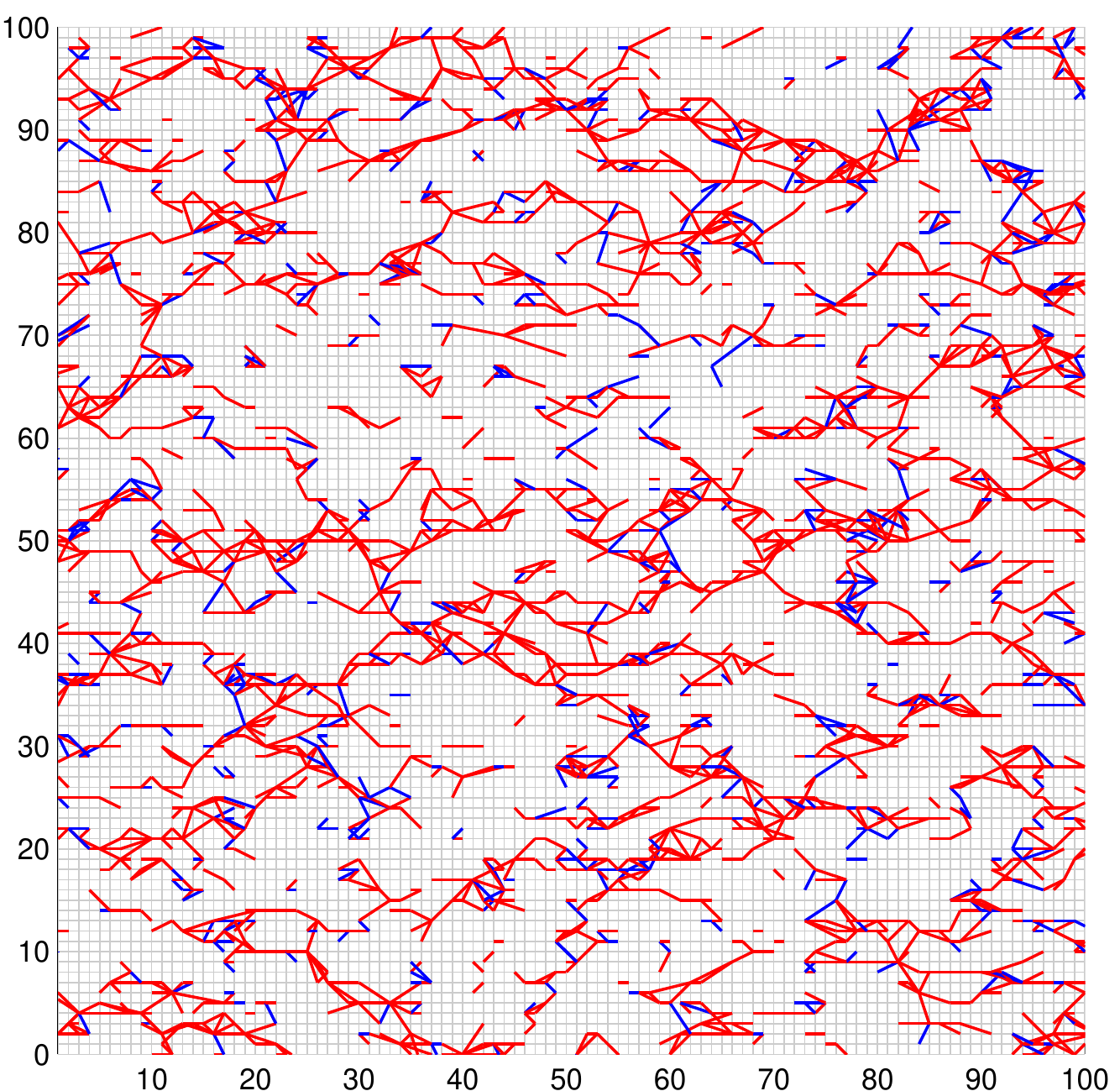}}
\put(0.5,0.52){\includegraphics[width=0.5\unitlength,angle=0]{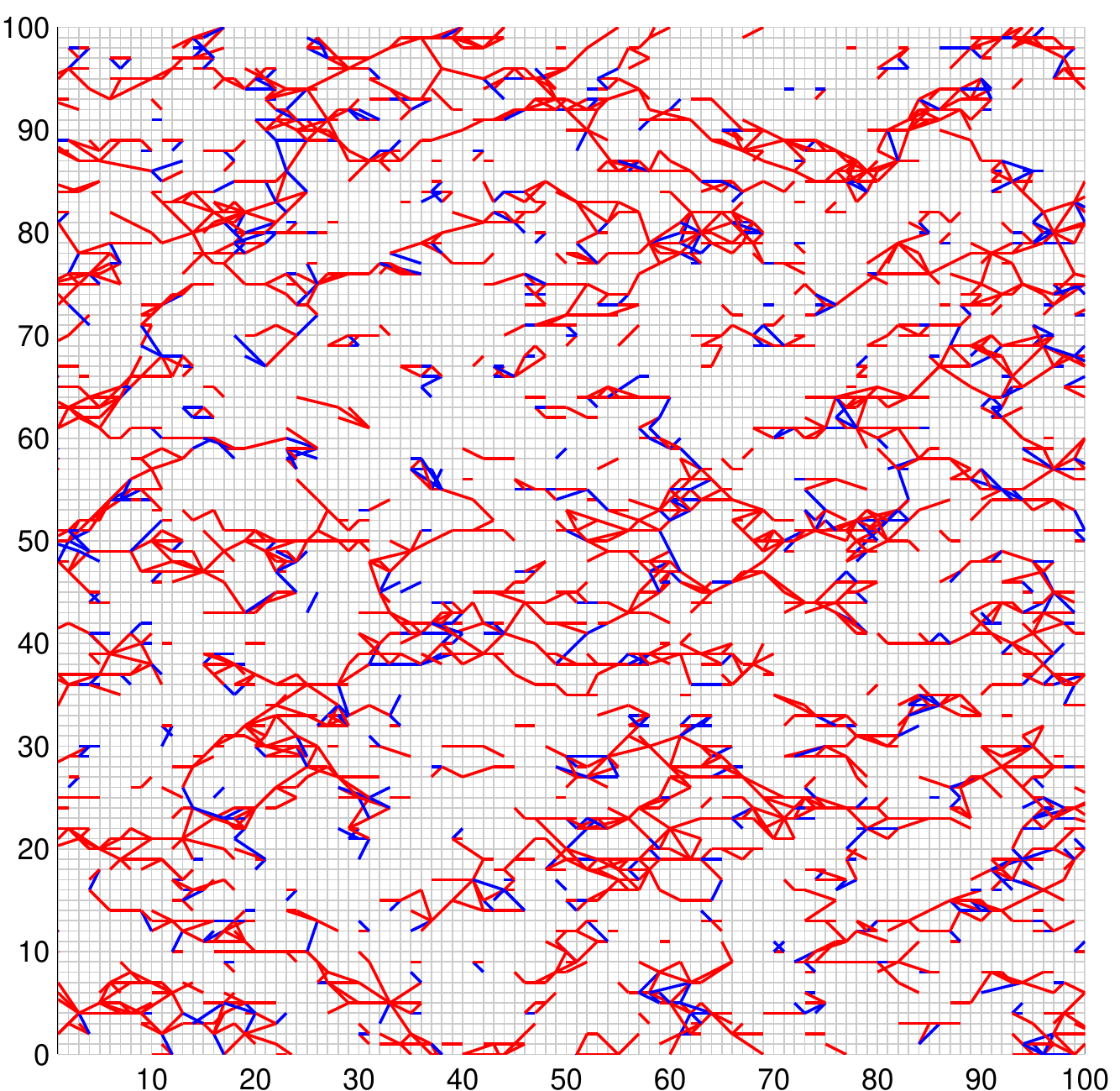}}
\put(0,0.02){\includegraphics[width=0.5\unitlength,angle=0]{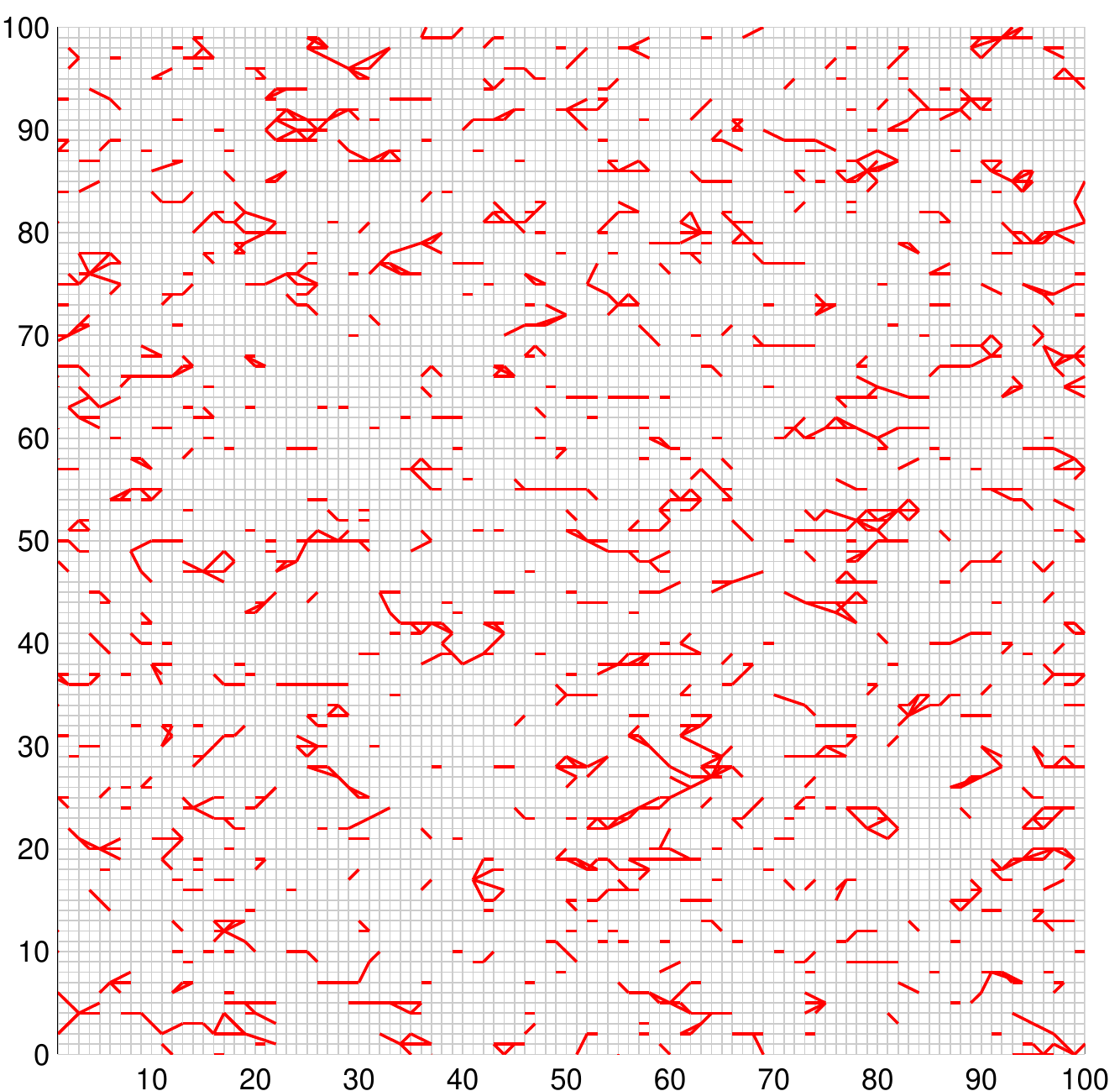}}
\put(0.5,0.02){\includegraphics[width=0.5\unitlength,angle=0]{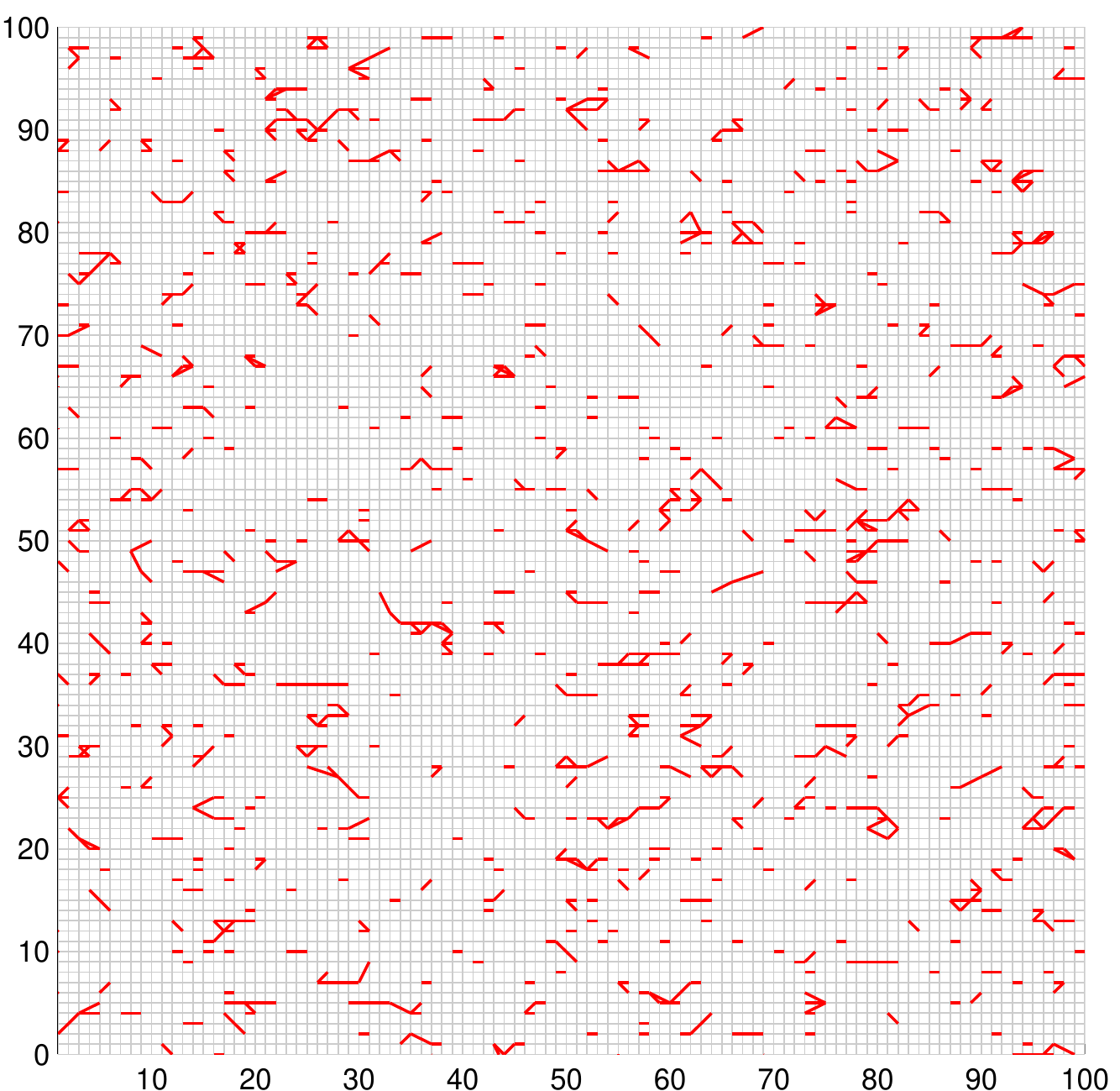}}
\put(0.22,0.5){$\omega=0$}
\put(0.72,0.5){$\omega=10^{-3}$}
\put(0.22,0){$\omega=10^{-1}$}
\put(0.72,0){$\omega=1$}
\end{picture}
\caption{\label{fig:aktDo}Links in the set $L_S$ for $D=0.8$ and $\omega= 0, 10^{-3}, 10^{-1}, 1$. 
}
\end{figure}

We look at the temperature $T=0.05$ and simulate $2\cdot10^8$ jumps at
each frequency.  We plot each link in the set $L^{S(0.8)}$, that is,
all active links with $S$ chosen such that $D=0.8$. Links in blue
have $s_l<0$ while those in red have $s_l>0$. Figure
\ref{fig:aktDo} shows such activity maps for different frequencies. 
We can see that there is considerable overlap
between the maps, but also many links which differ. At high
frequencies, there is a tendency for smaller clusters to form, as
expected.

We observe that as $\omega$ increases there are less and less links
with $s_l<0$ (blue color), and for the highest frequencies there are no
links which have a negative $s_l$.  We can understand this by the fact
that the percolation cluster in DC is a complex structure where the
current in many places has to go opposite of the overall current
direction in order for the current to pass around difficult
regions. As the frequency increases, the need for this goes away and
links with negative $s_l$ become less important. We also find that the
values of the largest $s_l$ increase for large frequencies, and that
the value of $S(D)$ for fixed $D=0.8$ also increases. 

We can also see that the total number of links with $s_l>S(D)$ (the
size $|L^{S(D)}|$ of the set $L^{S(D)}$) decreases with increasing
frequency. However the number of links with nonzero $s_l$ (the size
$|L^{0}|$ of the set $L^0$) does not change significantly with the
frequency.

\begin{table}
\begin{tabular}{ccccccccc}
\hline
$\omega$&0&$10^{-5}$&$10^{-4}$&$10^{-3}$&$10^{-2}$&$10^{-1}$&1\\
\hline
$|L^0|$&59365&65921&65796&65891&65728&65612&65878\\
\hline
$|L^{S(0.8)}|$&3493&3412&3581&3255&2466&1477&1079\\
\hline
\end{tabular}
\caption{The number $|L^0|$ of links with nonzero $s_l$ as well as the
  number $|L^{S(0.8)}|$ of active links with $D=0.8$ for different $\omega$. }
\end{table}

Let us compare the AC maps with the DC
map which shows the full percolation network.
\begin{figure}
\setlength{\unitlength}{\textwidth}
\begin{picture}(1,1)(0,0)
\put(0,0.52){\includegraphics[width=0.5\unitlength,angle=0]{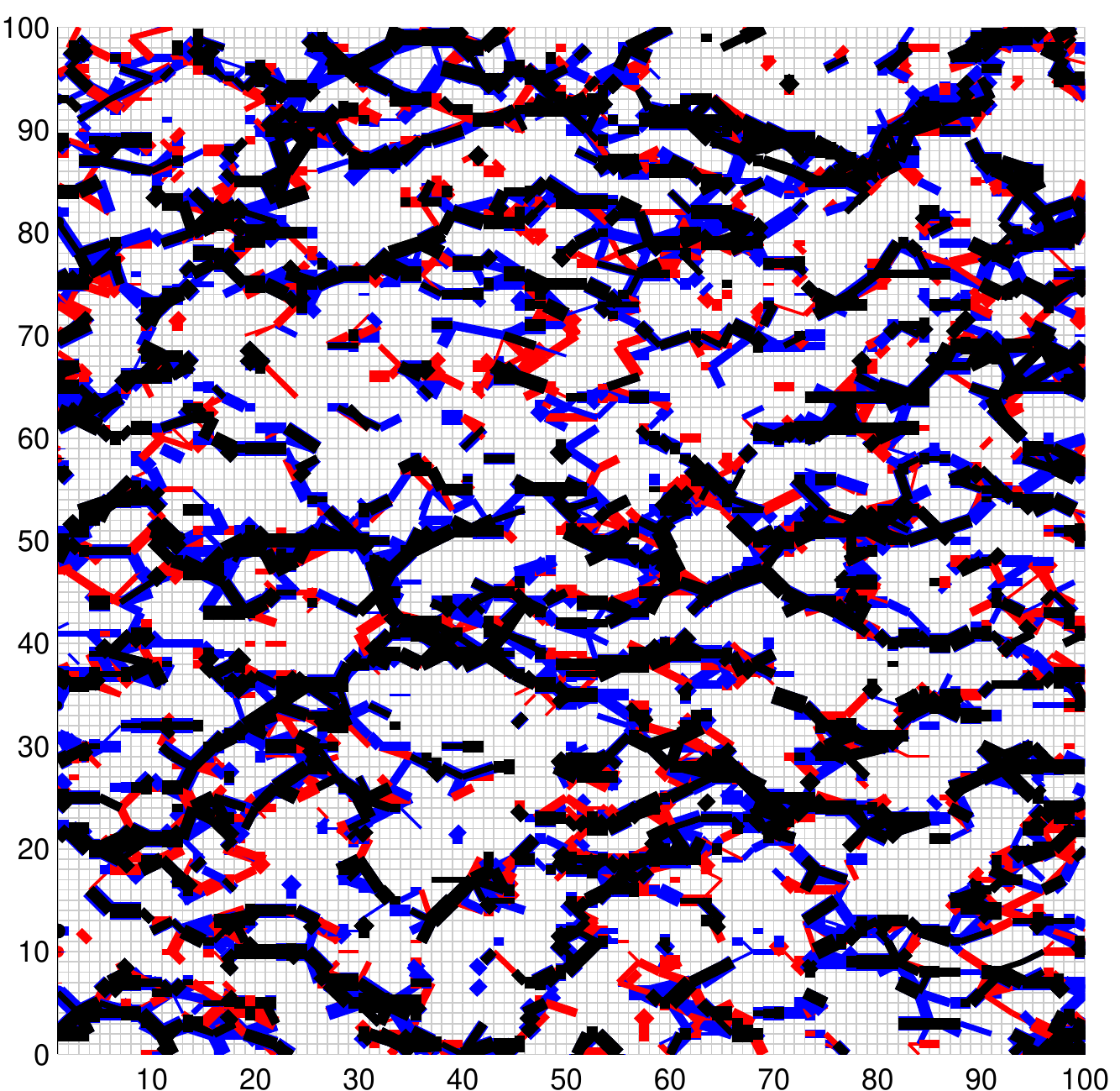}}
\put(0.5,0.52){\includegraphics[width=0.5\unitlength,angle=0]{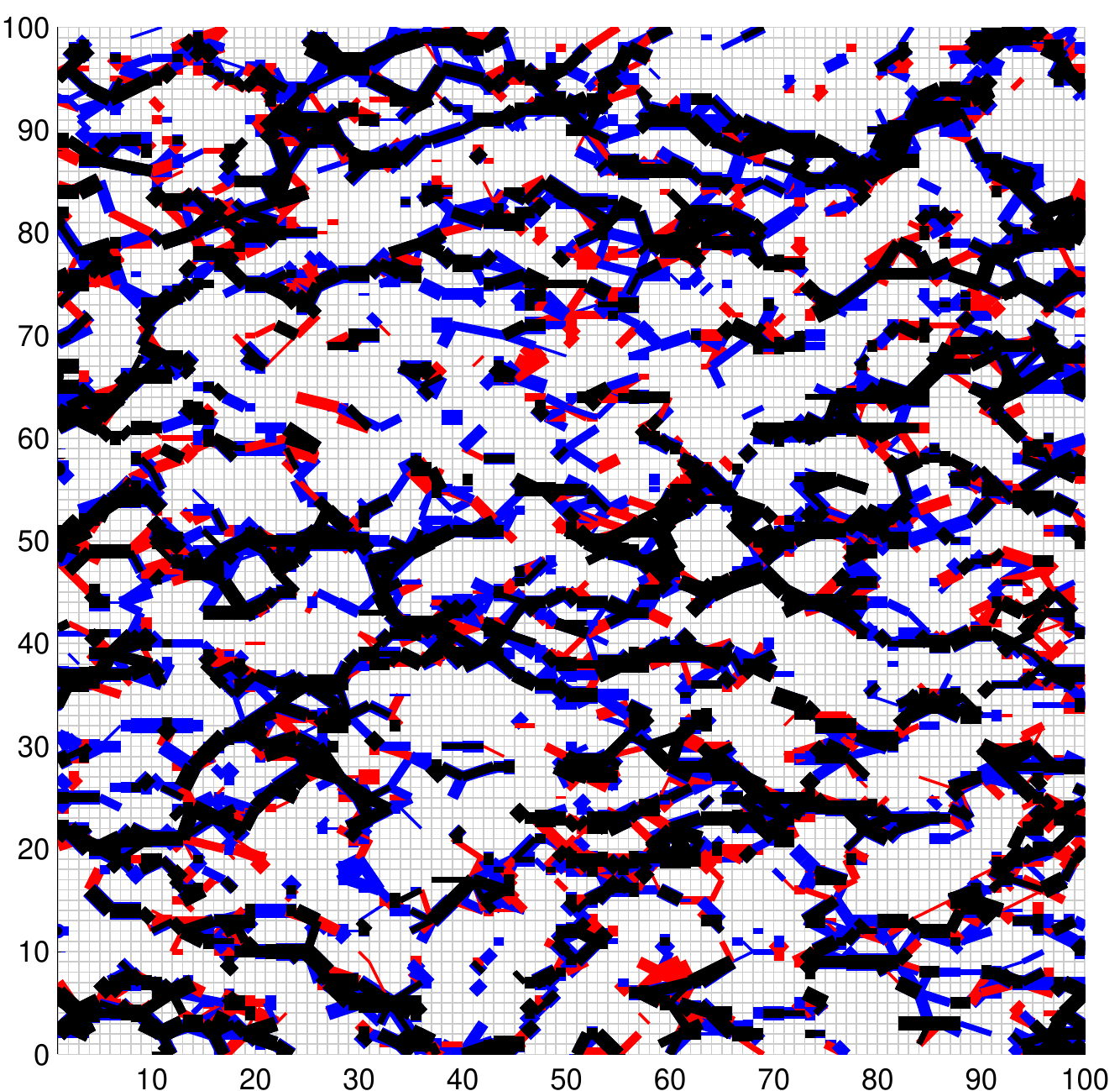}}
\put(0,0.02){\includegraphics[width=0.5\unitlength,angle=0]{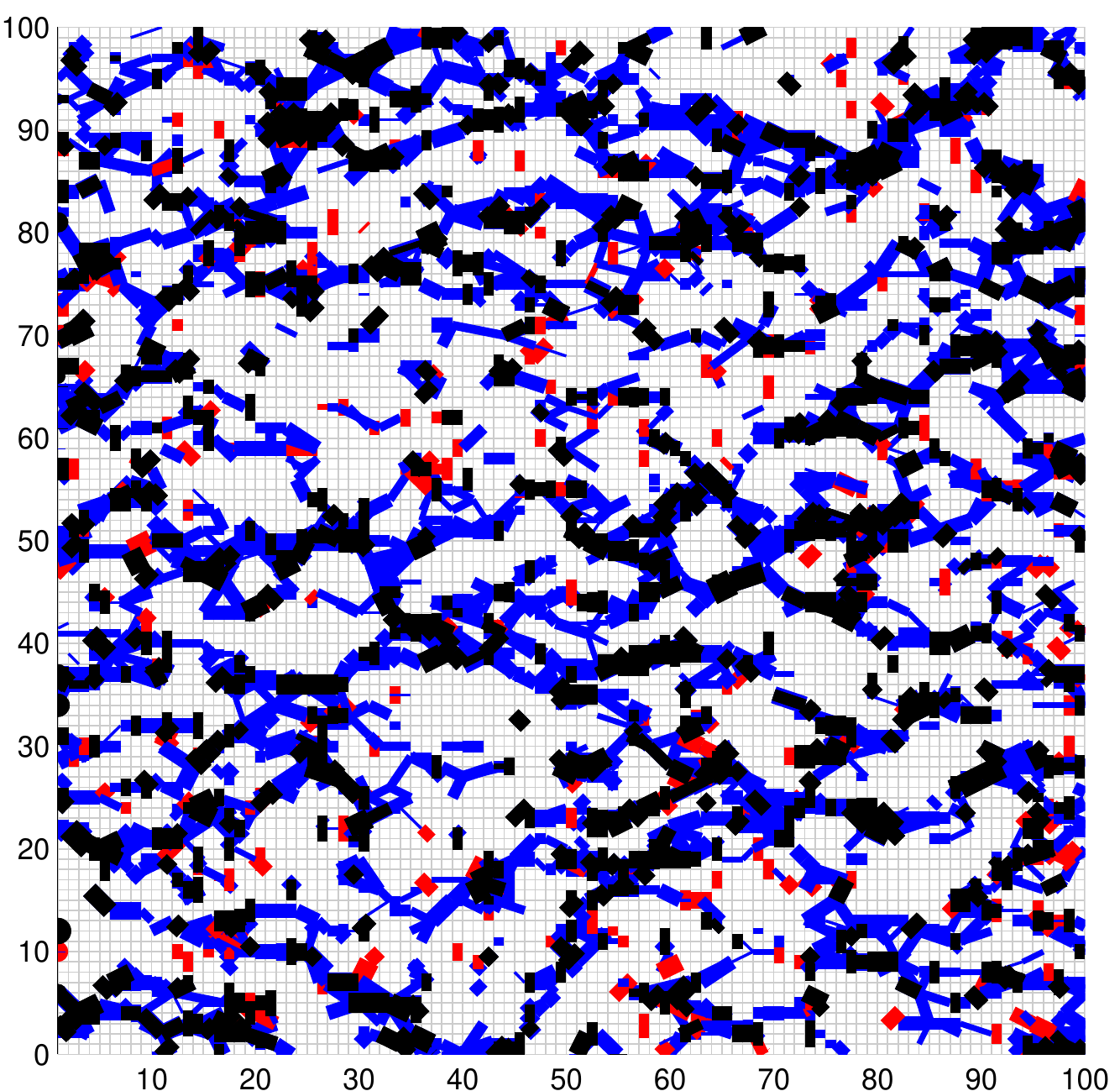}}
\put(0.5,0.02){\includegraphics[width=0.5\unitlength,angle=0]{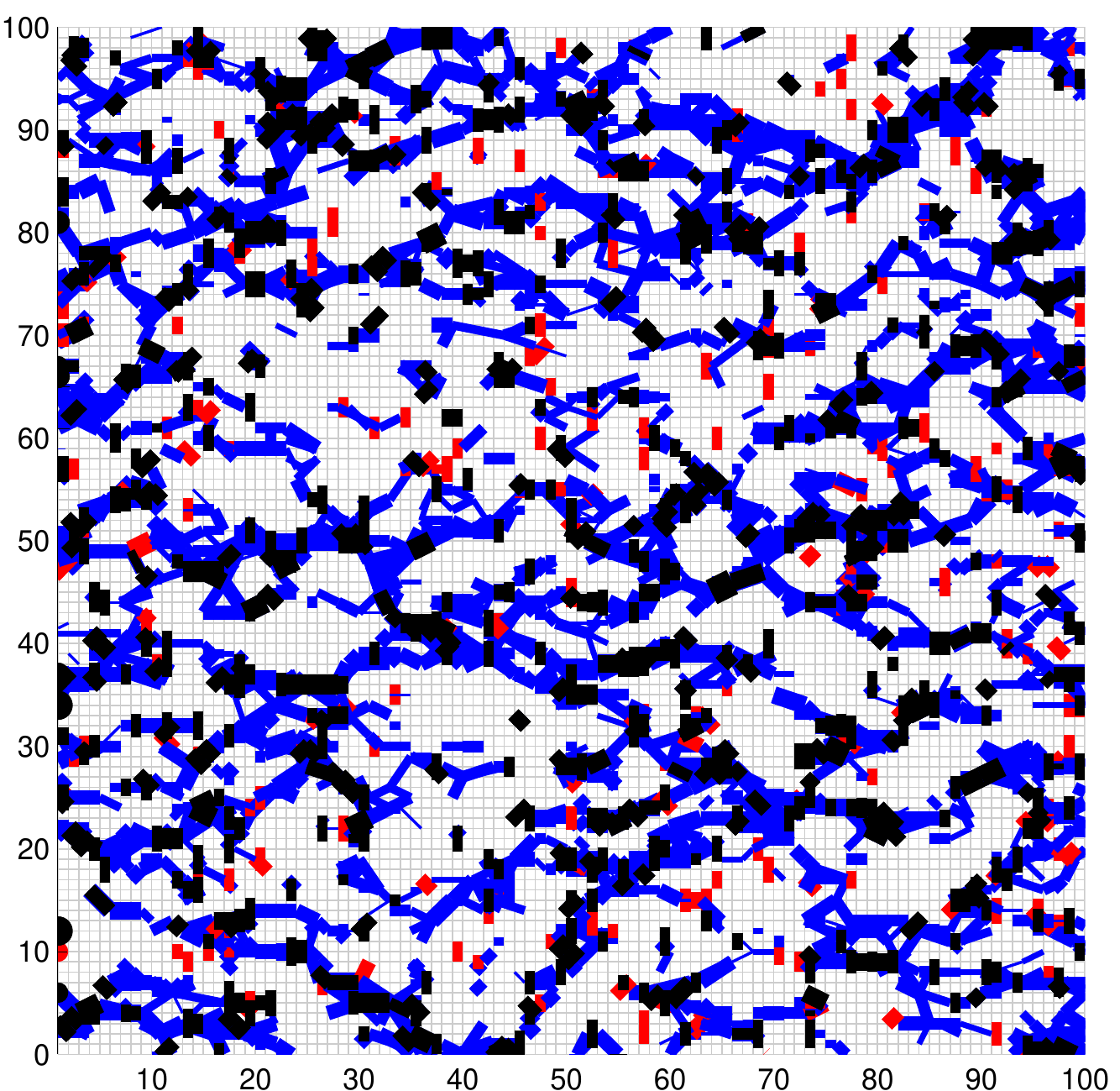}}
\put(0.22,0.5){$\omega=10^{-5}$}
\put(0.72,0.5){$\omega=10^{-3}$}
\put(0.22,0){$\omega=10^{-1}$}
\put(0.72,0){$\omega=1$}\end{picture}
\caption{\label{fig:aktComp}Comparison of the sets $L_S$ for $D=0.8$
  and $\omega=10^{-5}, 10^{-3}, 10^{-1}, 1$ with the
  corresponding DC set. Black links are present in both sets, while
  red only in the AC and blue only in the DC set.
}
\end{figure} 
We plot in Figure \ref{fig:aktComp} the comparison. Links which are
active in both the DC and AC maps are shown in black. Links in  blue are
only active in the DC map and those in red only in the AC map.  The width of
the lines are given by the formula
%\begin{equation}
%  \text{Line width} = 
$\lceil\ln(|s_l|-S(D)+1)\rceil$
%\end{equation}
so that lines with large $s_l$ are given larger width, but only
logarithmically so. For the black lines, the thickness is determined
from the value of $s_l$ in the AC map. We can see that at low
frequencies, when the conductivity is close to the DC value, the
structure is similar to the DC one. Most of the links which are not
present in both maps (red or blue) are thin lines which shows that
most of the difference comes from small fluctuations in the numbers
and does not significantly change the structure.  As the frequency
becomes so high that the conductivity starts to increase, we observe
that there are parts of the percolation cluster that are no longer
active. At the highest frequencies, there also appear isolated pairs
of sites outside of the percolation cluster which give large
contributions. However, it seems that even at $\omega=1$ there are
still clusters of more than two sites on the percolation cluster which
are active, while there hardly at any frequency are apparent isolated
clusters of more than two active sites. Also, at high frequencies
there are more black than red links, which means that most of the
transport still resides on the percolation cluster.

\begin{theacknowledgments}
JB thanks M. Palassini for hospitality and useful discussions. 
\end{theacknowledgments}

\bibliographystyle{aipproc}   % if natbib is available

\end{document}